%% file: root.tex
\documentclass[a4paper, 10pt, conference]{ieeeconf}      

\IEEEoverridecommandlockouts                              

\input{preamble}

\title{\LARGE \bf
Algorithmic design and implementation considerations of deep MPC
}

\author{Prabhat K. Mishra, Mateus V. Gasparino, and Girish Chowdhary
	\thanks{P. K. Mishra is with the Department of Artificial Intelligence at Indian Institute of Technology, Kharagpur, India. (\tt{pkmishra@ai.iitkgp.ac.in}).}
    \thanks{M. V. Gasparino is with the University of Illinois at Urbana Champaign (UIUC), IL, USA. The publication was written prior to M. V. Gasparino joining Amazon.}
		\thanks{G. Chowdhary is with the University of Illinois at Urbana Champaign (UIUC), IL, USA (\tt{girishc@illinois.edu}).}
	}

\begin{document}

\maketitle
\thispagestyle{empty}
\pagestyle{empty}

\begin{abstract}
Deep Model Predictive Control (Deep MPC) is an evolving field that integrates model predictive control and deep learning. This manuscript is focused on a particular approach, which employs deep neural network in the loop with MPC. This class of approaches distributes control authority between a neural network and an MPC controller, in such a way that the neural network learns the model uncertainties while the MPC handles constraints. The approach is appealing because training data collected while the system is in operation can be used to fine-tune the neural network, and MPC prevents unsafe behavior during those learning transients. This manuscript explains implementation challenges of Deep MPC, algorithmic way to distribute control authority and argues that a poor choice in distributing control authority may lead to poor performance. A reason of poor performance is explained through a numerical experiment on a four-wheeled skid-steer dynamics.  
\end{abstract}

\section{INTRODUCTION}
Deep learning-based MPC utilizes the constraint satisfaction capability of MPC and the function approximation capability of \emph{Deep Neural Network} (DNN). 
It is evident that machine learning and deep learning techniques can enhance the performance of MPC. However, there is still a lack of successful results demonstrating the use of deep learning to augment MPC policies in uncertain environments, employing deep neural networks (DNNs) to learn uncertainties online \cite{2023_survey_automotive}. One of the results to address such problem was reported in \cite{2025_deepMPC}. 

\begin{figure}
	\centering
	\begin{adjustbox}{width =\columnwidth}
	\includegraphics{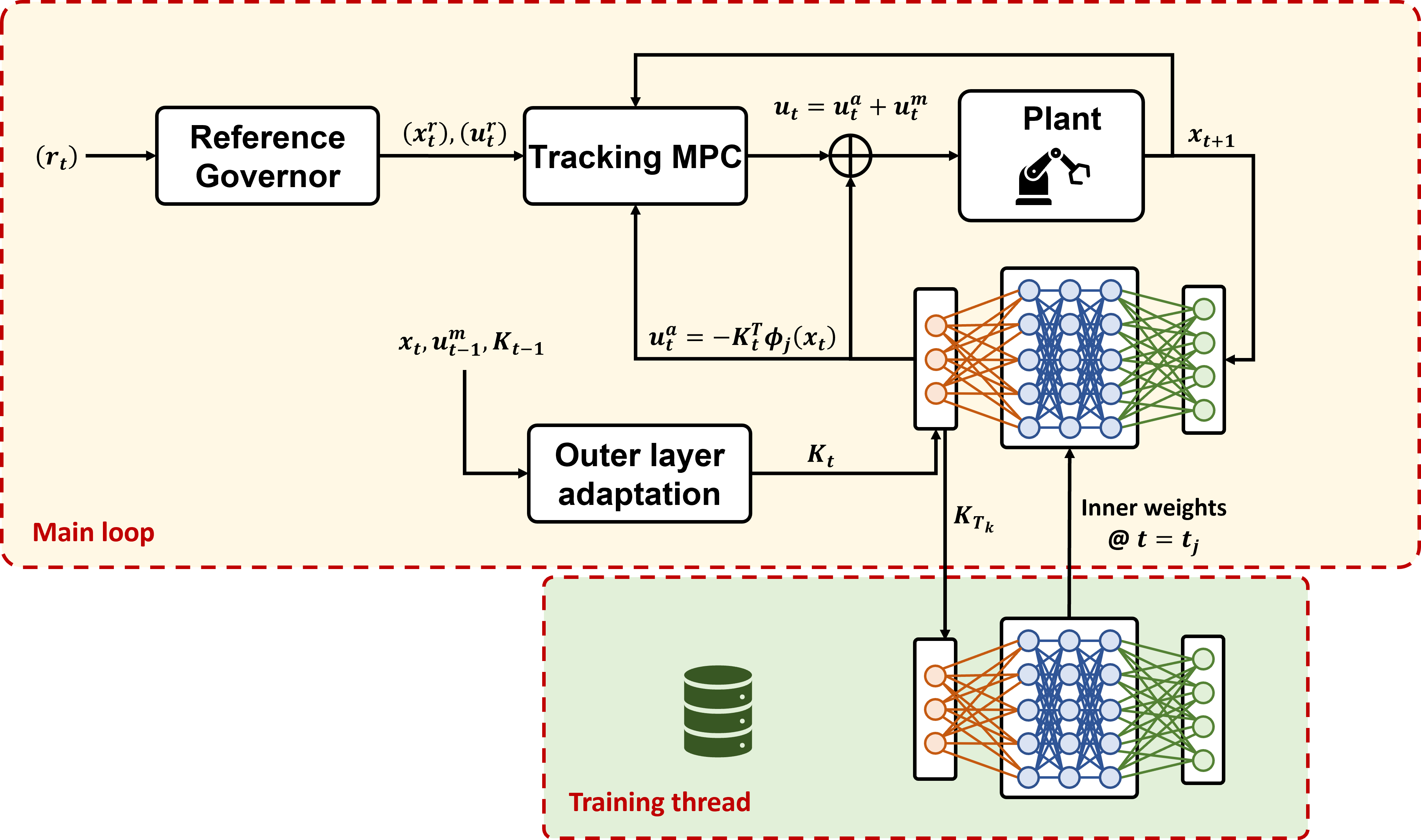}		
	\end{adjustbox}
	  \caption{Deep MPC has a neural network in the loop with MPC in which the outermost layer is adapted at each time instant and its output is stored in a replay buffer to create training dataset to train its inner layers intermittently in a remote training thread.}
	  \label{fig:deep_mpc_block}
\end{figure}

Deep MPC \cite{2025_deepMPC} is based on Tube-MPC \cite{Mayne_tube_NLMPC} and inherits mathematical analysis from tube-MPC itself, which guarantees input-to-state stability of the controlled system. In addition, the idea of \cite{2025_deepMPC} is based on collecting the training data for DNN online while the system is in operation and the training data is collected from a DNN itself, similar to \cite{DMRAC}. This mechanism is summarized in Fig. \ref{fig:deep_mpc_block} and will be discussed in more detail later.

To integrate a DNN within the tube-MPC framework, it is essential that the DNN produces bounded outputs. This allows MPC to treat the DNN output as an additional bounded disturbance \cite{NNMPC_Findeisen}. This requirement can be addressed by using bounded activation functions in the output layer and constraining the weights (trainable parameters of the outermost layer) through projection onto a bounded set. While it may seem that bounding the weights of the output layer is necessary solely because tube-MPC is designed to handle only bounded disturbances (i.e., the DNN outputs), this constraint is also crucial for mitigating the parameter drift phenomenon commonly encountered in adaptive control. 

Parameter drift phenomenon also reminds the crash of X-15 in October 1967 in which the pilot Michael Adams died and due to the incident flight testing of adaptive control stopped for around the next thirty years \cite{2010_adaptive_book_Hovakimyan}. However, several new methods of robust adaptive control have also been developed later. One of the approaches to avoid the parameter drift phenomenon is based on projecting the learned weights on a bounded set. 

In the context of Deep MPC \cite{2025_deepMPC}, the output of a DNN itself is used to train another DNN. Therefore, the output dataset of a DNN with projected weights and bounded activation function results in vanishing gradient phenomenon during the training. 
As a result, in many cases performance of tube-MPC and deep MPC either remains the same or in some cases the underlying optimization problem of deep MPC becomes infeasible. In particular, let $u_t := u_t^a + u_t^m$ be the control applied at time $t$, where $u_t^a$ is the learning component and $u_t^m$ is the MPC component. Since $u_t$ should be bounded for all $t$, a careful distribution of control authority between $u_t^a$ and $u_t^m$ is required. When $u_t^a$ does not get enough control authority, learning may not happen and therefore, their performance may be the same. When $u_t^m$ does not get enough control authority, the underlying optimization problem can become infeasible.
This manuscript provides a tutorial introduction to Deep MPC, its algorithmic details and provides numerical justification about distributing the control authority between $u_t^a$ and $u_t^m$.

Our notations are standard and mostly defined before their first use. We describe the problem setup in \S \ref{s:setup} and provide the overall architecture of deep MPC in \S \ref{s:deep_MPC}. The implementation details, algorithms and explanations are provided in \S \ref{s:implementation}. The reason of poor learning in deep MPC is explained through a numerical experiment in \S \ref{s:experiment} and finally we conclude in \S \ref{s:conclusion}. 

\section{Problem setup}\label{s:setup}
We are interested in a dynamics, which has the form 
\begin{equation}\label{e:dynamics}
x_{t+1} = f(x_t) + g(x_t) \left( u_t + h(x_t)\right),
\end{equation}
where 
\begin{enumerate}[leftmargin = *, nosep, label=(1-\alph*), widest = b]
\item $x_t \in \mathcal{X} \subset \mathbb{R}^d$, $u_t \in \mathcal{U} \subset \mathbb{R}^m$ are state and action, respectively, at time $t$.
\item The term $g(x_t)h(x_t)$ represents state-dependent disturbance and there exists $w_{\max} > 0 $ such that $\norm{g(x)h(x)} \leq w_{\max}$ for all $x \in \mathcal{X}$.
\item The functions $f$ and $g$ are known and Lipschitz continuous.
\item The function $h$ is unknown but continuous \footnote{The term $h(x)$ is considered state-dependent matched uncertainty at the state $x$. Refer to \cite{Pavon_unmatched} for the definitions of matched and unmatched uncertainties.},
\item $\mathcal{X}$ and $\mathcal{U}$ are compact sets.
\item The matrix $g(x)$ is left invertible for each $x \in \mathcal{X}$.
\end{enumerate}
There are many practical applications where an uncertain system follows the recursion \eqref{e:dynamics}. In a practical continuous-time system such as mobile robots, digital controllers can be implemented by either discretizing the dynamics before computing the control or discretizing the controller before implementing it. Optimal control formulations such as reinforcement learning and MPC are generally based on discretized dynamics while adaptive control implementations are generally based on discretized controller.

In some cases we may be interested in a control constraint of the form $\abs{u_t^{i}} \leq u_{\max}^{i}$ for $i = 1, \ldots, m$. These constraints can be written in the form $\norm{u_t}_{\infty} \leq u_{\max}$ by scaling $g(x_t)$ \cite[Remark 1]{PDQ_Policy}. Therefore, for simplicity, we define $\mathcal{U} := \{u \in \mathbb{R}^m \mid \norm{u}_{\infty} \leq u_{\max} \}$. 

The deep MPC \cite{2025_deepMPC} control $u_t$ has two components namely a learning component $u_t^a$ and a MPC component $u_t^m$. The composite control $u_t = u_t^a + u_t^m$ is applied to the system.
In the problem setup, we have assumed $h:\mathcal{X} \rightarrow \mathbb{R}^d$ is continuous and $\mathcal{X}$ is compact. Therefore, $h(x)$ will be bounded for all $x\in \mathcal{X}$. Since $h(x_t)$ is additive to $u_t$, we can think about designing $u_t^a = -h(x_t)$, if possible. Since $h$ is unknown, we are interested in learning $u_t^a$. The term $u_t^m$ should be used to fulfil other control objectives such as regulating states from a given state to a desired state. Since the control authority is limited by the constraint $\norm{u_t}_{\infty} \leq u_{\max}$, we need to redistribute the control authority $u_{\max}$ between $u_t^a$ and $u_t^m$. 

In the next section, we will describe the overall deep MPC architecture, then focus on the implementation details, specially how the control authority can be distributed between $u_t^a$ and $u_t^m$ algorithmically.

\section{Deep MPC}\label{s:deep_MPC}
The overall deep MPC architecture consists of a reference governor, tracking MPC, adaptation mechanism and two neural networks as shown in Fig. \ref{fig:deep_mpc_block}. Both neural networks have the same architecture with one linear output layer followed by a layer with bounded activation functions such as Tanh, sigmoidal, etc. The remaining architecture of the neural networks is flexible and the choice depends on the type of problems. For simplicity, we can consider fully connected layers too.  

Reference governor and tracking MPC modules are same as tube MPC \cite{Mayne_tube_NLMPC} with modifications required to adjust the adaptation mechanism. In other words, to steer a dynamical system \eqref{e:dynamics} from the initial state $x_0$ to the target state $x^s$ while satisfying the state and action constraints described with \eqref{e:dynamics}, we are interested in designing the control $u_t = u_t^a + u_t^m$ at time $t$, where $u_t^a$ is the output of the neural network in the main loop; please see Fig. \ref{fig:deep_mpc_block}. Reference governor takes slowly varying reference $r_t$ as input and suggests a trajectory from $x_0$ to $r_t$. For simplicity, we consider $r_t = x^s = 0$, which means the reference governor is designed offline for a regulation problem. For a tracking problem, reference governor is invoked whenever $r_t$ changes \cite{Mayne_tube_NLMPC, ref:PSC18}. The trajectory generated by reference governor need to have sufficient margin from the boundaries of $\mathcal{X}$ so that the states of the uncertain system \eqref{e:dynamics} controlled by the tracking MPC stay in $\mathcal{X}$. Therefore, tightening of constraints before designing the reference governor is required, which will be discussed in the next section.

The adaptation mechanism as shown in Fig. \ref{fig:deep_mpc_block} updates the parameters of the output layer (wights and biases) by assuming the hidden layers as a fixed feature basis function. Since the ideal feature basis function is not known in case of unstructured uncertainties, we utilize the second neural network. To train this second neural network, we use the data available in the replay buffer. To create the replay buffer data, we use output of neural network (main loop) as label and corresponding state as input with some experience selection criterion \cite{experience_selection}. We can intermittently (at time instants $(t_j)_{j \in \mathbb{N}}$) replace the parameters of the hidden layers of neural network (main loop) by the parameters of the hidden layers of the second neural network.

\section{Implementation details of Deep MPC}\label{s:implementation}
We have assumed that the state-dependent additive disturbance $\norm{g(x)h(x)} \leq w_{\max}$ for all $x \in \mathcal{X}$. We can consider the set $\mathcal{X}$ as our intended region of operation.
Once we have a discrete dynamics \eqref{e:dynamics}, the next step to implement deep MPC is the computation of some acceptable value of $w_{\max}$. 

Most of the commercial devices come with their ratings. Therefore an acceptable value of $w_{\max}$ can be safely prescribed. For example, the maximum crosswind speed for an air-plane is prescribed in its flying handbook \cite[Chapter 9]{1972_handbook_federal}. Related discussion about the Boeing aircraft is provided in the report \cite{safety_boeing}. When such bounds are not provided by the manufacturer, specially when the device is assembled in the laboratory, experiments can be performed to compute bounds similar to the approach available in \cite[Fig. 6]{2021_robust_tumbling}. As an example consider that 36 kts is provided as the maximum crosswind speed as company policy for landing but it can also vary according to the runway conditions.  
Consider a case as another example where a drone is tasked with delivering packages in an urban environment. Suppose initial experiments reveal that the position estimation error varies between 1 and 5 meters due to GPS signal interference in different urban settings. After consulting with domain experts, if the control engineer finds that the signal can be further degraded by certain environmental features like tall buildings or dense foliage, etc., they can prescribe a larger error bound (say 6 meters) and design the controller accordingly. 
For more algorithmic clarity and generalization of the approach, we can use data generation methods as discussed in \cite[Chapter 2]{2022_billard}. Suppose we have $M$ trajectories containing $(x_t,u_t)$ pairs and $T_i$ denotes the length of $i^{\text{th}}$ trajectory. We can follow the Algorithm \ref{algo:wmax_umax} to compute $w_{\max}$ and $u_{\max}^a$, where the variable $u_{\max}^a$ will be discussed in the context of learning in more detail and for our own understanding, we can call it a learning authority, i.\ e.\ $\norm{u_t^a}_{\infty} \leq u_{\max}^a$ for all $t$.   
\begin{algorithm}
\caption{Computation of bounds $w_{\max}, u_{\max}^a $}
\label{algo:wmax_umax}
\begin{algorithmic}[1]
    \State \textbf{Input:} $(x_j, u_j)_{j=0}^{T_i}$ for $i=1, \ldots, M$, $f$, $g$
    \State \textbf{Output:} $w_{\max}, u_{\max}^a$
    \State $w_{\max}, u_{\max}^a \gets 0$
    \For{$i \gets 1$ to $M$}
       \For{$j \gets 0$ to $T_i - 1$}
          \State $ w \gets x_{j+1} - f(x_j) - g(x_j)u_j$ 
          \State $w^a \gets g(x_j)^\dagger w$
          \State $w_{\max} \gets \max \{ w_{\max} , w \} $
          \State $u_{\max}^a \gets \max\{ u_{\max}^a, w^a \} $ 
       \EndFor
    \EndFor
    \State \textbf{return} $w_{\max}, u_{\max}^a$
\end{algorithmic}
\end{algorithm}

Now we need to design a reference governor. Reference governor provides a trackable reference trajectory, which can be tracked by the nominal dynamics. To clear the nomenclature, we can call \eqref{e:dynamics} our uncertain dynamics because it has an unknown term $h(x_t)$. The nominal dynamics is given by $x_{t+1} = f(x_t) + g(x_t)u_t^m =: \bar{f}(x_t, u_t^m)$. We can also use other names for the nominal dynamics such as simulator dynamics or the dynamics of the digital twin. In a regulation problem, if our set-point is $(x^s, u^s)$ such that $x^s = f(x^s) + g(x^s) u^s $ and the starting point $x_0$ is given, our control objective is to reach the set-point without violating the state and action constraints while minimizing some cost. We emphasize the definition of set-points for beginners because in most of the cases $u^s = 0$ may not serve the purpose. For example, in order to maintain the temperature of a room at certain level $x^s$, a certain amount of heating $u^s$ is required. Role of reference governor is to suggest a trajectory from $x_0$ to $x_s$ such that we can find $u_t \in \mathcal{U}$, which can maintain the states of the uncertain system \eqref{e:dynamics} in the operational region $\mathcal{X}$. Since the reference governor works on the nominal dynamics, we tighten the sets $\mathcal{X}$ and $\mathcal{U}$. This approach is called constraint tightening and is an active area of research in MPC \cite{2022_constraint_tightening, Mesbah_constraint_tightening, Muller_constraint_tightening, 2025_constraint_tightening}. For linear dynamical systems, the tightened constraint set $\mathcal{X}_r$ can be obtained by various types of robust positively invariant (RPI) sets by using different methods such as subtracting the evolution of the disturbance set from the set $\mathcal{X}$. These methods provide analytic expression of the tightened sets but they involve polytopic set computations and are therefore, computationally expensive. In practical implementations generally some approximation of RPI set or tightened set is used. For non-linear dynamical systems, constraint tightening is more difficult not only computationally but also analytically. 

As discussed in \cite[\S 7]{Mayne_tube_NLMPC}, constraints of the system \eqref{e:dynamics} are satisfied by the tightening the constraints of the nominal dynamics $\bar{f}$ used in the design of the reference governor and then forcing the states and control of the online MPC in a tube. The online MPC has constraints only on control, so the satisfaction of state constraints solely depend on the constraint tightening. Theoretical results guarantee that the states of the controlled system lie in a tube defined by level sets, which is defined in terms of the value function of the online MPC \cite[Proposition 1]{2025_deepMPC}.

Constraint tightening plays a very important role in tube-based MPC formulation and the intuition is that when the nominal dynamics satisfy the tightened constraints, the uncertain dynamics will satisfy the constraints given in \eqref{e:dynamics}. Since deep MPC has to distribute control authority between MPC component and learning component, we define $\bar{\mathcal{U}} := \{u \in \mathbb{R}^m \mid \norm{u}_{\infty} \leq u_{\max} - u_{\max}^a \}$ and compute the tightened control set $\mathcal{U}_r \subset \bar{\mathcal{U}}$ according to \cite[\S 7]{Mayne_tube_NLMPC}.  

Once we have tightened sets $\mathcal{X}_r \subset \mathcal{X}$ and $\mathcal{U}_r \subset \bar{\mathcal{U}}$, we can solve a finite horizon problem by using the nominal dynamics $\bar{f}(x_t, u_t^m) = f(x_t) + g(x_t)u_t^m$ as follows:
\begin{equation}\label{e:reference_governor}
\begin{aligned}
	\min_{(u_i^r)_{i=0}^{N-1}} & \quad \sum_{i=0}^{N-1} \norm{x_i^r}^2_Q + \norm{u_i^r}_R^2  \\
	\sbjto & \quad x_0^r = x_0, x_N^r = \zeros, \\
	& \quad x_{i+1}^r = \bar{f}(x_i^r, u_i^r) \in \mathcal{X}_r \subset \mathcal{X},  \\
	& \quad x_i^r \in \mathcal{U}_r \subset \bar{\mathcal{U}}; i = 0, \ldots, N-1,  
\end{aligned}
\end{equation}
to generate a reference trajectory $(x_t^r,u_t^r)_{t=0}^N$,  where $N$ is the optimization horizon. The design of reference governor is popular for designing hierarchical MPC to satisfy constraints \cite{2025_safe_kogel}. We can use CasADi \cite{casadi} to design a reference governor by using the algorithm \ref{algo:ref_governor}.

\begin{algorithm}
\caption{Reference governor}
\begin{algorithmic}[1]
    \State \textbf{Input:} $N, Q, R, x_0, x^s, u^s, \mathcal{X}_r, \mathcal{U}_r, \bar{f}$
    \State \textbf{Output:} $(x_t^r,u_t^r)_{t=0}^N$
    \State $x_0 = x_0 - x^s$
    \State Solve \eqref{e:reference_governor}
    \State $x_t^r \gets x_t^r + x^s$, $u_t^r \gets u_t^r + u^s$
    \State \textbf{return} $(x_t^r,u_t^r)_{t=0}^N$
\end{algorithmic}
\label{algo:ref_governor}
\end{algorithm}

The MPC component $u_t^m$ in Deep MPC is aware of the learning component $u_t^a$, therefore we need to design $u_t^a$ before $u_t^m$. As depicted in Fig. \ref{fig:deep_mpc_block} that the outermost layer of the neural network is updated at each time instant by keeping the inner layers fixed. The deep MPC approach allows the training of inner layers intermittently. Let $k^{\text{th}}$ training of inner layers starts at the time instant $T_k$ and $j^{\text{th}}$ training finishes before the time instant $t_j$.
We initialize $K_0$ such that $\norm{K_0^{(i)}} \leq \bar{W}_i$; $i = 1, \ldots , m$. For a given learning rate $0< \theta < 1$ and for $t \in \{t_j +1,  \ldots , t_{j+1} \} $, we employ the following weight update law:
\begin{equation}\label{e:update_law}
		\bar{K}_{t} = K_{t-1} + \frac{\theta}{\norm{\phi_j(\st_{t-1})}^2} \phi_j(\st_{t-1})\left( g(\st_{t-1})^{\dagger}(\st_{t} - \bar{f}(\st_{t-1}, \control_{t-1}^m)) \right)\transp , 
\end{equation}
where $g(\st_{t-1})^{\dagger} = \left( g(\st_{t-1})\transp g(\st_{t-1}) \right)^{-1}g(\st_{t-1})\transp $ represents the pseudo-inverse of the left invertible matrix $g(\st_{t-1})$. The above weight update law has been derived in \cite{haddad_adaptive} for structured uncertainties and has been used in \cite{MWGC} along with MPC. Notice that the first element in $\phi_j(\cdot)$ is one. Therefore, $\norm{\phi_j(x)}^2 \geq 1$ for all $x \in \R^d$ and $j \in \Nz$, which avoids any possibility of division by zero. We employ the discrete projection method to ensure boundedness of $K_t^{(i)}$ for $i= 1, \ldots , m$, as follows:
\begin{equation}\label{e:projection}
		K_t^{(i)} = \proj \bar{K}_t^{(i)} = \begin{cases} \bar{K}_t^{(i)} & \text{ if } \norm{\bar{K}_t^{(i)}} \leq \bar{W}_{i} \\
			\frac{\bar{W}_{i}}{\norm{\bar{K}_t^{(i)}}} \bar{K}_t^{(i)} & \text{ otherwise. } \end{cases}
\end{equation}

\begin{algorithm}
\caption{Transfer learning in the neural network for $t \in [t_j, t_{j+1}[$}
\label{algo:transfer_learning}
\begin{algorithmic}[1]
    \State \textbf{Input:} $x_t, x_{t-1}, u_{t-1}^m, \bar{f}, \phi_j, \bar{W}_i$ for $i=1, \ldots, m$
    \State \textbf{Output:} $K_t$ \Comment{Weights of the output layer}
    \State update \eqref{e:update_law}
    \State projection \eqref{e:projection}
    \State \textbf{return} $K_t$
\end{algorithmic}
\end{algorithm}

The Algorithm \ref{algo:transfer_learning} provides $K_t$ for $t\in \{ t_j+1, \ldots, t_{j+1} \}$ and is used to generate the learning component $u_t^a = -K_t^\top \phi_j(x_t)$ for $t\in \{ t_j, \ldots, t_{j+1} -1\}$. The pair $(x_t, u_t^a)$ are stored in a replay buffer for the supervised fine tuning of the neural network on the basis of some experience selection criterion \cite{experience_selection, svd_cl}. The Algorithm \ref{algo:transfer_learning} has also been used in \cite{2023_unmatched}.

Once all these modules are ready, we can create the main module of deep MPC. In order to design the online reference tracking MPC, we first choose symmetric positive definite matrices $Q, R \succ \zeros$, which can be different from those chosen for the reference governor. Let 
 $ \costps(\st_{t+i \mid t}, \control_{t+i \mid t}) \Let \norm{\st_{t+i \mid t} - \st_{t+i}^r}_Q^2 + \norm{\control_{t+i \mid t} - \control_{t+i}^r}_R^2$
be the cost per stage at time $t+i$ predicted at time $t$ and let $\costfinal(x) \Let x\transp Q_f x$ be the terminal cost with $Q_f \succ 0$. 

Let us define 
\begin{equation}\label{e:cost_function}
V_m(x_{t\mid t}, (u_{t+i\mid t})_{i=0}^{N-1}) \Let c_f(\st_{t+N \mid t}) + \sum_{i=0}^{N-1} c_s(\st_{t+i \mid t}, u_{t+i \mid t}) .
\end{equation}  
The online reference tracking MPC minimizes \eqref{e:cost_function} at each time instant $t$ under the following constraints:
\begin{align}
&x_{t \mid t} = x_t \label{e:constraint_initial}\\
& u_{t \mid t} + u_t^a \in \mathcal{U} \label{e:constraint_first_control}\\
& x_{t+ i +1 \mid t} = f(x_{t+i \mid t}) + g(x_{t+i \mid t})u_{t+i\mid t} \text{ for } i = 0, \cdots, N-1   \label{e:constraint_dynamics}\\
& u_{t + i\mid t} \in \bar{\mathcal{U}} \text{ for } i = 1, \cdots, N-1 . \label{e:constraint_remaining_control}
\end{align}
Notice that the constraint \eqref{e:constraint_first_control} is different from the constraints present in tube-MPC formulation \cite{Mayne_tube_NLMPC}. We define the underlying optimal control problem 
\begin{equation}\label{e:MPC}
V_m(\st_{t}) \Let \min_{(\control_{t+i \mid t})_{i=0}^{N-1}} \left\{ \eqref{e:cost_function} \mid \eqref{e:constraint_initial}, \eqref{e:constraint_first_control}, \eqref{e:constraint_dynamics}, \eqref{e:constraint_remaining_control} \right\}.
\end{equation}

The overall deep MPC algorithm is given in Algorithm \ref{algo:deep_MPC}.
\begin{algorithm}
\caption{Deep MPC}
\label{algo:deep_MPC}
\begin{algorithmic}[1]
    \State \textbf{Input:} $\mathcal{X}, \mathcal{U}, x^s, u^s, \{ t_j\} , \{ T_k \} $, replay buffer 
    \State \textbf{Output:} $u_t$
    \State Compute $w_{\max}, u_{\max}^a$ by Algorithm \ref{algo:wmax_umax}
    \State Constraint tightening according to \cite[\S 7]{Mayne_tube_NLMPC}
    \State Generate the reference trajectory $(x_t^r, u_t^r)$ by Algorithm \ref{algo:ref_governor}
    \For{each $t$}
     \State measure state $x_t$
     \If{$t=T_k$}
        \State activate the training thread
        \State take random samples from the replay buffer
        \State freeze output layer weights to be $K_{T_k}$
        \State train the network 
        \State $k \leftarrow k+1$
    \EndIf
    \State Compute $K_t$ by Algorithm \ref{algo:transfer_learning}
    \If{$t=t_j$}
    \State update $\phi_j$ in the main loop
    \State $j \leftarrow j+1$
    \EndIf
    \State $u_t^a = -K_t^\top \phi_j(x_t)$
    \If{($x_t, u_t^a$) satisfies the experience selection criterion}  
    \State Store ($x_t, u_t^a$) in the replay buffer
    \EndIf
    \State Solve \eqref{e:MPC} to get $u_t^m$
    \State \textbf{return} $u_t = u_t^a + u_t^m$
    \State $t \leftarrow t+1$
    \EndFor    
\end{algorithmic}
\end{algorithm}

\section{Numerical experiment}\label{s:experiment}\label{s:experiment}

We consider a continuous time simplified non-slip four-wheeled skid-steer dynamics of the robot,  which is used in  \cite{gasparino2023cropnav,higuti2019under, 2025_deepMPC} for robot navigation in an agricultural crop field. The states of the robots are represented by $s \Let \bmat{x& y& \theta& v& \omega}\transp$, where $x \in [-2, 1]$ is the robot's longitudinal position across the lane, $y \in [-0.5, 0.5] $ is the lateral position (cross-track error), $\theta \in [-\pi/2, \pi/2 ]$ is the heading angle with respect to the lane, $v \in [-1.5, 1.5]$ is the difference between the speed of the robot and the reference speed $v_{r}=0.8$, and $\omega \in [-3, 3]$ is the robot's angular rate. The control $u \Let \bmat{F_{L} & F_{R}}\transp$, where $F_{L}, F_{R} \in [-10, 10]$ are the left and right exerted forces used to move the robot's wheels. The continuous time dynamics is given by
\begin{equation}\label{e:skid_steer_robot}
    \bmat{\dot{x} \\ \dot{y} \\ \dot{\theta} \\ \dot{v} \\ \dot{\omega}} =
    \bmat{
         (v + v_r)  \cos \theta \\
        (v + v_r)  \sin \theta \\
        \omega \\
        0 \\
        0}
    + \bmat{
        0 & 0 \\
        0 & 0 \\
        0 & 0 \\
        \frac{1}{M}   & \frac{1}{M}   \\
       - \frac{b}{I}  &  \frac{b}{I}
    } \left( \bmat{F_L \\ F_R} - R \right),
    \end{equation}
where the constants $M=15$, $I=0.1$, and $b=0.1$ are the robot's physical properties that correspond to the mass, moment of inertia, and distance between wheels. The continuous time dynamics is discretized by using the sampling time $T_s = 0.05$ seconds to get the dynamics of the form \eqref{e:dynamics}.

The matched uncertainty $h(s) \Let -R$ represents unknown rolling resistance forces. For the purpose of simulation, we have generated $ R $ by using the expression $R^{(1)} = -2 \cos \theta + \omega y - v \theta + \omega$ and $R^{(2)} = 2(1- \sin \theta ) + x v -y -0.5y^2$ as in \cite{2025_deepMPC}.
The control objective is to steer the states of robot $s_t$ from $s_0 =  \bmat{-1& -0.25& \pi/4& 0& -\pi/8}\transp$ to the equilibrium $s_e = \bmat{0& 0& 0& -v_r& 0}$. 

\begin{figure}
	\centering
	\begin{adjustbox}{width =\columnwidth}
	\includegraphics{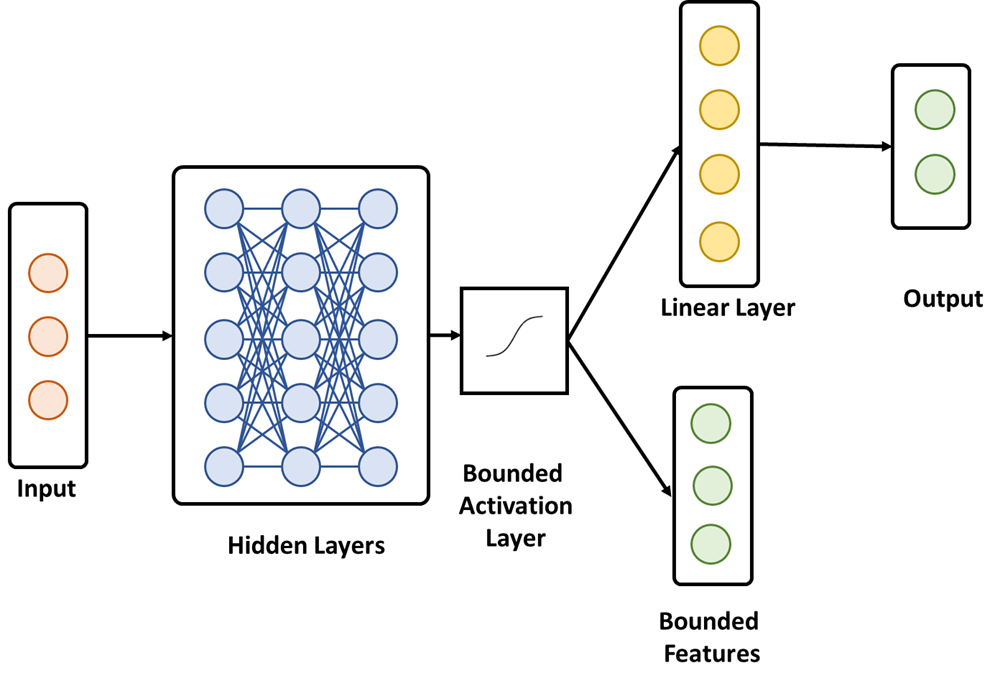}		
	\end{adjustbox}
	  \caption{The neural network architecture used in Deep MPC.}
	  \label{fig:NN}
\end{figure}

To design the deep MPC, we used a neural network with three hidden layers of sizes $\{ 8, 12, n_2 \}$, where $n_2=4$ represents the outermost layer. We used $tanh$ as a bounded activation function in the last hidden layer and Rectified Linear Unit (ReLU) in the remaining hidden layers; see Fig. \ref{fig:NN}. The weights of the linear output layer are updated by using \eqref{e:update_law} and \eqref{e:projection} with learning rate $\theta = 0.5$ and bounds $\bar{W}_i$ in \eqref{e:projection} is obtained by 
\begin{equation}\label{e:projection_bound}
\bar{W}_i = \frac{u_{\max}^a}{\sqrt{n_u(1+0.25 n_2)}} \text{ for } i = 1, \ldots, n_u + 1.
\end{equation}
The hidden layers of Deep MPC are trained at each $20$ time steps using SGD with learning rate 0.01. 
We choose the optimization horizon $N=10$, the state penalty matrix $Q$ as a diagonal matrix with entries $\{ 0.5, 2, 1, 0.5, 5 \}$, the control penalty matrix $R=I$ and the terminal cost matrix $Q_f = 10^5 I$. We have implemented the reference governor \eqref{e:reference_governor} by using the same parameters $Q,R$ as the online MPC, $N=100$ and heuristically tightened the constraints by multiplying them with a factor of $0.9$. 
We follow the approach of \cite{svd_cl} for the inclusion and removal of data pairs based on the singular value maximization criterion. We carried out simulations by using a MATLAB-based software package MPCTools \cite{mpctools}.

\begin{figure}
	\centering
	\begin{adjustbox}{clip, trim=0.7cm 0.1cm 1.1cm 0.7cm,width =\columnwidth}
	\includegraphics{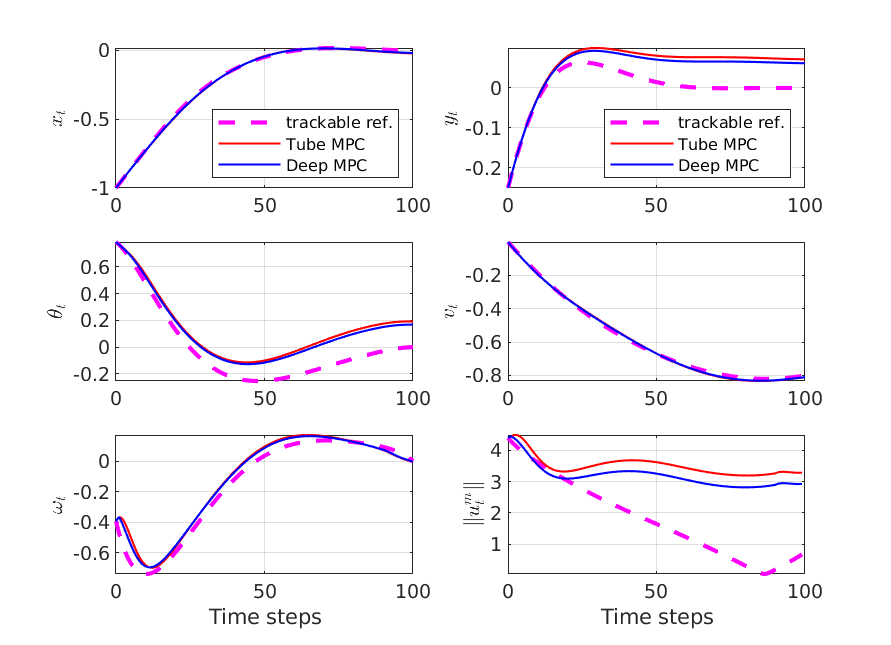}		
	\end{adjustbox}
	  \caption{In the absence of enough learning authority, performance of Deep MPC is close to that of tube MPC.}
	  \label{fig:skid_steer_robot_states}
\end{figure}

\begin{figure}
	\centering
	\begin{adjustbox}{clip, trim=0.7cm 0.1cm 1.1cm 0.7cm,width =\columnwidth}
	\includegraphics{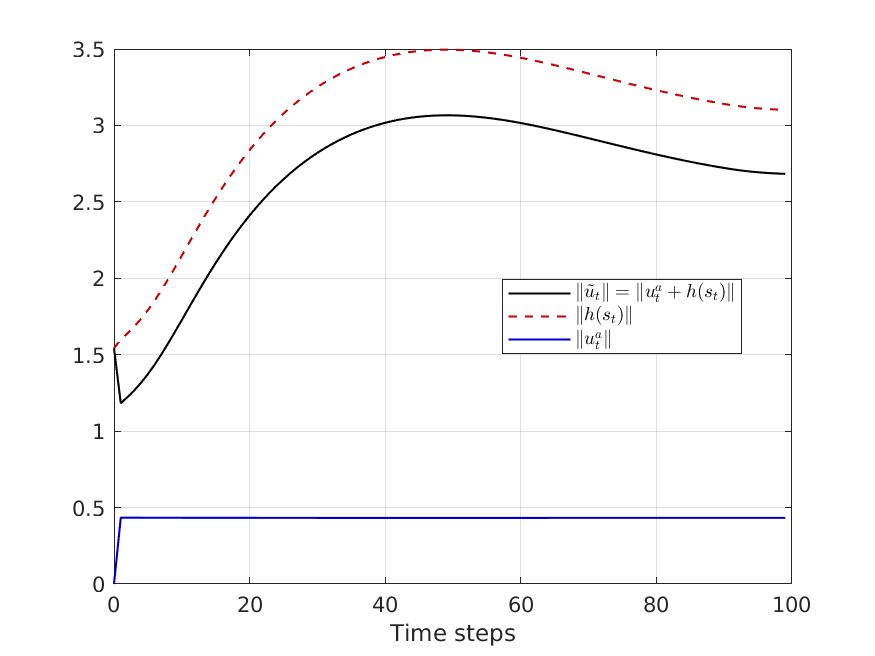}		
	\end{adjustbox}
	\caption{In the absence of enough learning authority, $u^a$ become saturated and therefore, $\tilde{u} = h+u^a$ changes according to $h$.}
	\label{fig:apparent_disturbance}
\end{figure}

In this section, we have chosen $u_{\max}^a = 0.6$ and experimental results are demonstrated in Fig.\ \ref{fig:skid_steer_robot_states} and Fig. \ref{fig:apparent_disturbance} for $100$ time steps. Fig.\ \ref{fig:skid_steer_robot_states} demonstrates that the trajectory of tube-MPC and deep MPC are very close to each other due to the absence of enough learning authority. The term $\tilde{u}_t =  u_t^a + h(s_t)$ in Fig. \ref{fig:apparent_disturbance} does not converge to some neighbourhood of origin instead it follows the trajectory of $h(s)$ with some offset. This phenomenon is due to the saturation of the learning control $u_t^a$ as shown in Fig. \ref{fig:apparent_disturbance}. This experiment concludes that learning is not happening due to the absence of enough learning authority and saturation of the training data $u_t^a$ resulting in the lack of relationship between $x_t$ and $u_t^a$. For practical purposes, we can use $u_{\max}^a$ little larger or equal to the value obtained by using Algorithm \ref{algo:wmax_umax}, similar to the numerical experiment in \cite{2025_deepMPC}.  

We considered the same numerical experiment as in \cite{2025_deepMPC} but we chose smaller $u_{\max}^a$ and demonstrated that deep MPC performs as tube-MPC (Fig. \ref{fig:skid_steer_robot_states}). We explained the reason of not getting advantage over tube-MPC in Fig. \ref{fig:apparent_disturbance}. The simulation result reveals that smaller $u_{\max}^a$ leads to saturation of $u_t^a$ and therefore learning does not happen. 

\section{CONCLUSIONS}\label{s:conclusion}
We have discussed about algorithmic implementation details of deep MPC. We have demonstrated through a numerical experiment that the poor choice of learning authority leads to poor performance. Future work may involve further simplification of constraint tightening or different approaches to satisfy state constraints. Evaluation of deep MPC for more challenging problems with different neural network architectures and experience selection criterion can be an interesting direction. The deep MPC algorithm is not suitable for random uncertainties but its integration with stochastic MPC approaches \cite{2016_review_SMPC, 2024_PCT_Mishra} can also be interesting.

\section*{ACKNOWLEDGMENT}
The first author is supported by the grant ANRF/ECRG/2024/004853/ENS.

\bibliographystyle{IEEEtran}
\bibliography{MPC_Learning}

\end{document}

%% file: preamble.tex
\usepackage{multibib}
\usepackage{graphicx}
\usepackage{mathptmx} 
\usepackage{newtxtext}
\usepackage{textcomp}
\usepackage[varg,bigdelims]{newtxmath}
\let\labelindent\relax
\usepackage{enumitem}

\usepackage{amsmath}
\usepackage{dsfont, url}
\usepackage{mathtools}
\usepackage[usenames, dvipsnames]{xcolor}
\usepackage[%
colorlinks={true},%
citecolor={Blue},%
urlcolor={PineGreen},%
linkcolor={Sepia}%
]{hyperref}
\usepackage{float}
\usepackage{newfloat}
\usepackage[skip=1pt]{caption}

\usepackage{amsmath} 
\usepackage{algorithm}
\usepackage[noend]{algpseudocode}
\usepackage{pgf,tikz}
\usepackage{adjustbox}
\usepackage{verbatim} 
\usepackage{xargs}
\usepackage{cite}
\usepackage{multirow}
\usepackage{pifont}

\DeclareMathOperator*{\sbjto}{s.\ t.\ }

\DeclareMathOperator{\proj}{Proj}

\renewcommand{\leq}{\leqslant}

\renewcommand{\geq}{\geqslant}

\newcommand{\R}{\mathds{R}}
\newcommand{\Nz}{\mathds{N}_0}

\newcommand{\bmat}[1]{\begin{bmatrix}#1\end{bmatrix}}
\newcommand{\abs}[1]{\left|#1\right|}
\newcommand{\norm}[1]{\left\|#1\right\|}

\renewcommand{\transp}{^\top}

\newcommand{\zeros}{\mathbf{0}}

\newcommand{\st}{x}

\newcommand{\control}{u}

\newcommand{\costps}{c_{\mathrm{s}}}
\newcommand{\costfinal}{c_{\mathrm{f}}}

\newcommand{\Let}{\coloneqq}

\allowdisplaybreaks